\documentclass[eqsecnum,showpacs,pre,twocolumn]{revtex4}
\usepackage{graphicx}
\usepackage{amsmath}
\usepackage{bm}
\usepackage{amsfonts}
\usepackage{amssymb}
\usepackage[usenames]{color}%
\begin{document}
\title{Directed Percolation with a Conserved Field and the Depinning Transition}
\author{Hans-Karl Janssen}
\affiliation{Institut f\"{u}r Theoretische Physik III, Heinrich-Heine-Universit\"{a}t,
40225 D\"{u}sseldorf, Germany}
\author{Olaf Stenull}
\affiliation{Department of Physics and Astronomy, University of Pennsylvania, Philadelphia,
Pennsylvania 19104, USA}
\date{\today}

\begin{abstract}
Conserved directed-percolation (C-DP) and the depinning transition of a
disordered elastic interface belong to the same universality class as has been
proven very recently by Le Doussal and Wiese [Phys. Rev. Lett.~\textbf{114},
110601 (2015)] through a mapping of the field theory for C-DP onto that of the
quenched Edwards-Wilkinson model. Here, we present an alternative derivation of
the C-DP field theoretic functional, starting with the coherent state path
integral formulation of the C-DP and then applying the
Grassberger-transformation, that avoids the disadvantages of the so-called
Doi-shift. We revisit the aforementioned mapping with focus on a specific term
in the field theoretic functional that has been problematic in the past when it
came to assessing its relevance. We show that this term is redundant in the
sense of the renormalization group.
\end{abstract}
\maketitle

%{{\color{PineGreen}{#1}}}
%{{\color{ProcessBlue}{#1}}}
%{{\color{BurntOrange}{#1}}}

%{{\color{PineGreen}{#1}}}
%{{\color{ProcessBlue}{#1}}}
%{{\color{BurntOrange}{#1}}}

%\pacs{64.60.Ht, 05.40.-a, 05.70.Jk, 64.60.ah, 64.60.av, 64.60.De}

\section{Introduction}

About 20 years ago, van Wijland, Oerding, and Hilhorst \cite{WiOeHi98}
introduced a model of the propagation of an epidemic in a population of
fluctuating density. Healthy (inactive) and sick (active) individuals, also
called $A$ and $B$ particles, diffuse freely and independently on a lattice of
dimension $d$, and react as $A+B\rightarrow2B$, $B\rightarrow A$, therefore
holding the total number of particles globally constant. As long as the
diffusion constants of both particle types are non-zero, a consistently
renormalizable field theory, commonly referred to as \emph{directed
percolation with a conserved field} (DP-C), can be derived, and the universal
scaling properties can be calculated in an $\varepsilon$-expansion
\cite{OeWijLerHi00,Ja01,JaTa05}. The model features a continuous, a
tricritical \cite{KrSchSch89}, and a fluctuation induced discontinuous
transition \cite{OeWijLerHi00} depending on the ratio of the diffusion
constants. A variation of the model where only the active individuals (the
agent) can diffuse, has become to known as the \emph{conserved
directed-percolation} (C-DP) model. In contrast to DP-C, C-DP has proven to be
notoriously difficult when it comes to renormalized field theory~\cite{Wij02}.
Numerical studies, however, have been fruitful, and it has been established
that C-DP and certain sandpile-models belong to the same universality class,
the so called Manna-class of self-organized criticality, and it was argued
that, \textit{inter alia}, the depinning model of interfaces in random media
belongs to this class
too~\cite{Ma90,DiVeZa98,VeDiMuZa98,RoPaVe00,PaVe00,AlMu02,BoChaDoMu07,HeHiLue08}%
. Very recently, Le Doussal and Wiese (LeDW) \cite{DouWie15} proved the latter
by presenting an exact mapping of C-DP to the quenched Edwards-Wilkinson (qEW)
model that describes the depinning transition of a disordered elastic
interface \cite{NaFi92,NaFi93,NaStTaLe92,LeNaStTa97}.

In this paper, we revisit the work of LeDW and we shed light on the mapping
from C-DP to qEW from somewhat different angles. First, we re-derive the field
theoretic response functional for C-DP (the starting point of the mapping)
using the so-called Grassberger transformation, an approach that, as we think,
is more natural than that taken in previous derivations in that the actual
number densities of particles serve as field variables and that avoids
problematic physical interpretations of imaginary fields. Then we discuss the
mapping in a way that is slightly different from that of LeDW with focus on a
specific term in the field theoretic functional. LeDW showed that this term
primarily changes the friction coefficient and they claimed that the ensuing
remaining term is \emph{irrelevant}. We show that this term is \emph{redundant}
in the sense of the renormalization group (RG) \cite{Weg74,ZinnJust89}, i.e.,
it does not need its own renormalization and hence it does not impact the
scaling behavior.

\section{Model and field theoretic functional}
The C-DP model is based on the reactions
\begin{subequations}
\label{Reaktionen}%
\begin{align}
A+B  &  \leftrightarrow2B\,,\label{Reakt1}\\
B  &  \rightarrow A\,. \label{Reakt2}%
\end{align}
Note that Eq.~(\ref{Reakt1}) includes both, forth and back reactions. The
$B$-particles diffuse in a $d$-dimensional volume whereas the $A$-particles
are immobile. It is well known, that the corresponding coherent state path
integral (CSPI)-action \cite{Doi76,GraSche80,Pel84,Uwe14,TaHoVo05,Wie15} is given by%
\end{subequations}
\begin{align}
\mathcal{S}  &  =\int d^{d}x\int_{-\infty}^{\infty}dt\Big\{(\hat{a}%
-1)\partial_{t}a+(\hat{b}-1)\partial_{t}b+D\nabla\hat{b}\cdot\nabla
b\nonumber\\
&  -k_{1}\big(\hat{b}^{2}-\hat{a}\hat{b}\big)ab-k_{2}\big(\hat{a}\hat{b}%
-\hat{b}^{2}\big)b^{2}-k_{3}\big(\hat{a}-\hat{b}\big)b\Big\}\,,
\label{CSPI-act}%
\end{align}
where $a,\hat{a}$ and $b,\hat{b}$ are the coherent fields describing the
species $A$ and $B$, respectively, that are subject to the initial and final
conditions $\hat{a}(\infty)=\hat{b}(\infty)=1$, $a(-\infty)=b(-\infty)=0$. It
is important to note that these fields are in general complex and do not
correspond to particle densities. These densities are given by $n_{A}=\hat
{a}a$ and and $n_{B}=\hat{b}b$ which are real and non-negative by construction.
To proceed from the CSPI-action (\ref{CSPI-act}) to a field theoretic
functional, previous studies~\cite{TaHoVo05,Wie15} have relied on the so-called
Doi-shift $\hat{a}=1+\tilde{a}$, $\hat{b}=1+\tilde{b}$. This approach has the
disadvantage that it produces, in a Langevin interpretation, mixed real and
imaginary noise, the latter resulting from the annihilation of $A$-particles in
reaction (\ref{Reakt1}). Furthermore, this approach has the disadvantage that
masks the conservation property of the total number of particles $A$ and $B$
since the coherent fields $a$ and $b$ are not particle densities. As we will
show, these problems can be avoided by switching to a description based on the
particle densities with help of the quasicanonical Grassberger-transformation
(see the Appendix for details and background information)
\begin{subequations}
\begin{align}
\hat{a}  &  =\exp(\tilde{n}_{A})\,,\quad a=n_{A}\exp(-\tilde{n}_{A})\,,\ \\
\hat{b}  &  =\exp(\tilde{n}_{B})\,,\quad b=n_{B}\exp(-\tilde{n}_{B})\,,
\end{align}
with $\tilde{n}_{A}(\infty)=\tilde{n}_{B}(\infty)=0$ and $n_{A}(-\infty
)=n_{B}(-\infty)=0$. When source fields $\rho_{A}$ and $\rho_{B}$ that feed
additional particles into the system are admitted, the resulting action is
\end{subequations}
\begin{align}
\mathcal{A}  &  =\int d^{d}x\int_{-\infty}^{\infty}dt\Big\{\tilde{n}%
_{A}\partial_{t}n_{A}+\tilde{n}_{B}\partial_{t}n_{B}\nonumber\\
&  +D\big(\nabla\tilde{n}_{B}\cdot\nabla n_{B}-n_{B}(\nabla\tilde{n}_{B}%
)^{2}\big)-\rho_{A}\tilde{n}_{A}-\rho_{B}\tilde{n}_{B}\nonumber\\
&  -\big(\exp(\tilde{n}_{B}-\tilde{n}_{A})-1\big)k_{1}n_{A}n_{B}\nonumber\\
&  -\big(\exp(\tilde{n}_{A}-\tilde{n}_{B})-1\big)\big(k_{2}n_{B}^{2}%
+k_{3}n_{B}\big)\Big\}\,. \label{action-2}%
\end{align}
This action guarantees the conservation of the total particle density
$n_{A}(\mathbf{x},t)+n_{B}(\mathbf{x},t)$, which can be seen as follows.
We demand the invariance of $\mathcal{A}$ under the symmetry
transformation
\begin{subequations}
\label{Conserv-Sym}%
\begin{align}
\tilde{n}_{A}(\mathbf{x},t)  &  \mapsto\tilde{n}_{A}(\mathbf{x},t)+\phi
(t)\,,\\
\tilde{n}_{B}(\mathbf{x},t)  &  \mapsto\tilde{n}_{B}(\mathbf{x},t)+\phi(t)
\end{align}
for any purely time-dependent function $\phi(t)$ with
$\phi(\infty)=0$. This symmetry transformation implies
that
\end{subequations}
\begin{align}
\frac{d}{dt}  &  \int d^{d}x\,\big(n_{A}(\mathbf{x},t)+n_{B}(\mathbf{x}%
,t)\big)\nonumber\\
&  =\int d^{d}x\big(\rho_{A}(\mathbf{x},t)+\rho_{B}(\mathbf{x},t)\big)\,,
\label{conservation}%
\end{align}
i.e., the average particle density is constant if particle sources are absent.
Note that the conservation-symmetry transformation Eqs.~(\ref{Conserv-Sym})
simply changes the density response fields with an additive contribution that
is linear in the generator $\phi$. In comparison, the corresponding
transformation in the formulation based on the Doi-shift is very clumsy. Hence,
also in this respect, the density variables are advantageous over the coherent
fields: they make transparent the role of the response field as the generators
of the particle-conservation symmetry.

Now we turn to the scaling behavior of the fields under coarse graining. The
fields $\tilde{n}_{A}$, $n_{A}$, $\tilde{n}_{B}$ and $n_{B}$ can be rescaled so
that all of them attain a positive scaling dimension (see Appendix A
and the argumentation below). Hence, we can truncate the expansion of the
exponentials in Eq.~(\ref{action-2}) after the quadratic term, and we can
neglect the irrelevant diffusional noise as well other irrelevant
contributions. After letting $n_{A}+n_{B}\rightarrow c$,
$\tilde{n}_{A}\rightarrow\tilde{c}$, $n_{B}\rightarrow n$, $\tilde{n}%
_{B}-\tilde{n}_{A}\rightarrow\tilde{n}$, $\rho_{A}+\rho_{B}\rightarrow\rho$,
and $\rho_{B}\rightarrow\sigma$ so that $n$ is now the particle density of the
agent and $\sigma$ is the corresponding source field, we obtain the dynamical
response functional~\cite{Ja76,DeDo76,Ja92}
\begin{align}
\mathcal{J}  &  =\int d^{d}x\int_{-\infty}^{\infty}dt\, \lambda\Big\{\tilde
{n}\lambda^{-1}\partial_{t}n-\tilde{n}\nabla^{2}n+\tilde{c}\lambda
^{-1}\partial_{t}c\nonumber\\
&  -\tilde{c}\nabla^{2}n-\rho\tilde{c}-\sigma\tilde{n} -\tilde{n}%
\big(gc-fn-\kappa\big)n-\frac{1}{2}\tilde{n}^{2}n\Big\} \, . \label{J-functional}%
\end{align}
For an alternate derivation of this functional based on the Doi-shift, see Appendix~\ref{appenTrunc}.

Note that $\mathcal{J}$ corresponds to the well known coarse grained effective Langevin
description (in Ito-interpretation and without the sources)
\begin{subequations}
\label{CDPLang}%
\begin{align}
\lambda^{-1}\partial_{t}n  &  =\nabla^{2}n+\big(gc-fn-\kappa\big)n+\eta\,,\\
\lambda^{-1}\partial_{t}c  &  =\nabla^{2}n\,,\\
\langle\eta(\mathbf{x},t)\rangle &  =0\,,\\
\langle\eta(\mathbf{x},t)\eta(\mathbf{x}^{\prime},t^{\prime})\rangle &
=\lambda n(\mathbf{x},t)\delta(\mathbf{x}-\mathbf{x}^{\prime})\delta
(t-t^{\prime})\,,
\end{align}
of C-DP. Note also that $f=g$ if the back reaction in (\ref{Reakt1}) is
forbidden. 

Dimensional analysis shows that the various quantities superficially
scale in terms of an appropriate inverse length-scale $\mu$ as $(\tilde
{n},\tilde{c},\kappa)\sim\mu^{2}$, $(n,c)\sim\mu^{d-2}$, $(f,g)\sim\mu^{4-d}$
if $\mathbf{x}\sim\mu^{-1}$ and $\lambda t\sim\mu^{-2}$. This signals that
$d_{c}=4$ is the upper critical dimension of the absorbing transition below
which both coupling constants $f$ and $g$ become relevant. It can be
easily shown that the response functional $\mathcal{J}$ encompasses all
relevant operators. All other operators, i.e., all other monomials that can be constructed from the fields $n,\tilde{n}$ and $c,\tilde{c}$ and their derivatives, are irrelevant since
their superficial scaling dimensions are larger than $d+2$, and the corresponding 
coupling constants have negative scaling dimensions lower than $-2$ near
$d=d_{c}=4$. In particular the diffusional noise $n(\nabla\tilde{n})^{2}$ is
irrelevant.

Alternatively, using $m=c-n$ instead of $c$, Eqs.~(\ref{CDPLang}) can be
recast as
\end{subequations}
\begin{subequations}
\begin{align}
\lambda^{-1}\partial_{t}n  &  =\nabla^{2}n-\lambda^{-1}\partial_{t}m\,,\\
\lambda^{-1}\partial_{t}m  &  =-\big(gm-(f-g)n-\kappa\big)n-\eta\,.
\end{align}
This form suggests that the time-integrated local agent density $\int dt\,n\sim
s$ is another potentially useful choice for the independent
density-field instead of $n$ itself. Indeed, this field is an
essential ingredient in the LeDW mapping of C-DP to qEW, see below.

In the following, we consider processes beginning at some time $t>0$. Hence,
$\sigma(\mathbf{x},t)$ and $n(\mathbf{x},t)$ are zero if $t\leq0$. We assume
that the inactive particles are placed into the system homogeneously with
density $c_{0}$ at some time $t_{0}<0$. Therefore $\rho(\mathbf{x}%
,t)=c_{0}\delta(t-t_{0})+\sigma(\mathbf{x},t)$, and $c(\mathbf{x}%
,t)=c_{0}\theta(t-t_{0})$ for $t<0$. Hence the time integral in
Eq.~(\ref{J-functional}) can be reduced to only positive times with an initial
condition $c(\mathbf{x},0)=c_{0}$.

To establish closer contact to the work by LeDW, we now let
$\tilde{c}+\tilde{n}\to\tilde{n}^{\prime}$, $\tilde{n}\to
g\tilde{\zeta}$, $m=c-n\to g^{-1}(\kappa-\zeta)$,
and obtain%
\end{subequations}
\begin{align}
\mathcal{J}=\int &  d^{d}x\int_{0}^{\infty}dt\,\lambda\Big\{\tilde{n}^{\prime
}\big(\lambda^{-1}\partial_{t}n-\nabla^{2}n-g^{-1}\lambda^{-1}\partial
_{t}\zeta\big)\nonumber\\
&  +gn\bigl[\tilde{\zeta}\big((\lambda gn)^{-1}\partial_{t}\zeta
+\zeta+(f-g)n\big)-\frac{g}{2}\tilde{\zeta}^{2}\bigr]\Big\}\,,
\label{J-functional 2}%
\end{align}
where $\zeta(\mathbf{x},0)=\kappa-gc_{0}$. This is the starting point of the
LeDW mapping from C-DP to qEW. In this representation of the response
functional, diffusional motion is formally separated from the local
fluctuations described by the fields $\zeta$ and $\tilde{\zeta}$.

\section{The Le-Doussal-Wiese mapping of C-DP to qEW}
The essential tool of the LeDW mapping is the switch from the local
agent-density $n$ to its the time-integrated version $s$ which can be viewed
as an interface-height. Taking the same route, we define the new field and its
conjugated response field by%
\begin{subequations}
\begin{align}
s(\mathbf{x},t)  &  =\lambda g\int_{0}^{t}dt^{\prime}\,n(\mathbf{x},t^{\prime
})\,,\\
\tilde{s}(\mathbf{x},t)  &  =-\frac{1}{\lambda g}\frac{\partial\tilde
{n}^{\prime}(\mathbf{x},t)}{\partial t}\,. \label{s-Var}%
\end{align}
Because we are approaching the C-DP transition from the active side where $n(t)>0$, it is guaranteed that $s$ is monotonically increasing in $t$, and it can
therefore be used as a local time variable $t\rightarrow t(s,\mathbf{x})$,
with increment $ds=\lambda gn\,dt$, as long as $\mathbf{x}$ is held constant.
The introduction of the new field transforms the response functional
(\ref{J-functional 2}) to%
\end{subequations}
\begin{align}
\label{Jagain}
\mathcal{J}=\int &  d^{d}x\Big\{\int_{0}^{\infty}dt\,\lambda\tilde
{s}\big(\lambda^{-1}\partial_{t}s-\nabla^{2}s-k-\zeta\big)\nonumber\\
&  +\int_{0}^{\infty}ds\,\bigl[\tilde{\zeta}\big(\partial_{s}\zeta
+\zeta\big)-\frac{g}{2}\tilde{\zeta}^{2}+\tilde{\zeta}\alpha\lambda^{-1}%
\dot{s}\bigr]\Big\}\,,
\end{align}
where $\alpha=f/g-1\geq0$, $k=gc_{0}-\kappa$, and $c_{0}$ now denotes the
density of all, active and inactive, particles initially. As functions of $s$
instead of $t$, the fields $\tilde{\zeta}$ and $\zeta$ appear in the path
integral with weight $\exp(-\mathcal{J})$ in Gaussian form, and describe an
Ornstein-Uhlenbeck-process. They easily are integrated out leading to the
reduced response functional~\cite{footnote}
\begin{align}
\mathcal{J}_{red}  &  =\int d^{d}x\bigg\{\int_{0}^{\infty}dt\,\lambda\tilde
{s}(t)^{\prime}\bigl[\lambda^{-1}\dot{s}(t)-\nabla^{2}s(t)-k\bigr]\nonumber\\
&  +\int_{0}^{\infty}dtdt^{\prime}\Bigl[\alpha\tilde{s}%
(t)G\bigl(s(t)-s(t^{\prime})\bigr)\dot{s}(t^{\prime})^{2}\nonumber\\
&  -\frac{\lambda^{2}g}{2}\tilde{s}(t)C\bigl(s(t)-s(t^{\prime})\bigr)\tilde
{s}(t^{\prime})\Bigr]\bigg\}\,, \label{J-red}%
\end{align}
where%
\begin{equation}
\label{propAndCorr}
G(s)=\theta(s)\exp(-s)\,,\quad C(s)=\frac{1}{2}\exp(-\left\vert s\right\vert
)\,.
\end{equation}
Here and in the following, we always disregard contributions of initial
disturbances since we are interested in the steady state behavior.

The term with the dimensionless coupling constant $\alpha$ that arises from
the back-reaction of (\ref{Reakt1}) warrants further discussion. LeDW neglect
this unpleasant term after arguing that it is irrelevant in the sense of the
RG, a reasoning that has to be taken with a grain of salt for this particular
term. We will show that this term is redundant instead of irrelevant in the
sense of the RG: it does not require an independent renormalization and can
therefore be neglected. To this end, we group terms with time-derivatives of
$s$ together and write (suppressing the $\mathbf{x}$-dependence for notational
convenience)%
\begin{align}
\dot{s}(t)+ &  \alpha\int_{0}^{t}dt^{\prime}G(s(t)-s(t^{\prime}))\dot
{s}(t^{\prime})^{2}\nonumber\\
&  =\int_{0}^{t}dt^{\prime}\dot{s}(t^{\prime})\bigl[\delta(t-t^{\prime
})+\alpha G(s(t)-s(t^{\prime}))\dot{s}(t^{\prime})\bigr]\nonumber\\
&  =\int_{0}^{s(t)}ds^{\prime}\bigl[\delta(s(t)-s^{\prime})+\alpha
G(s(t)-s^{\prime})\bigr]\dot{s}^{\prime}(s^{\prime})\nonumber\\
&  =:\int_{0}^{s(t)}ds^{\prime}B(s(t)-s^{\prime})\dot{s}^{\prime}(s^{\prime
})\,,\label{Bewegl}%
\end{align}
where $\dot{s}^{\prime}(\mathbf{x},s^{\prime})=\partial_{t}s(\mathbf{x}%
,t(s^{\prime},\mathbf{x}))$ as a shorthand notation. $t^{\prime}$ and
$s^{\prime}$ are connected by the definition (\ref{s-Var}). Next, we transform
the response field $\tilde{s}$ by letting
\begin{equation}
\label{sTildeTrans}
\tilde{s}(t)\rightarrow\int_{t}^{\infty}dt^{\prime}\,\tilde{s}(t^{\prime
})K(s(t^{\prime})-s(t))\dot{s}(t)\,,
\end{equation}
where
\begin{equation}
K(s)=\delta(s)-\alpha G((1+\alpha)s)=:(1+\alpha)D((1+\alpha)s)\,.\label{D(s)}%
\end{equation}
is the inverse kernel of $B$. Then, after the additional
rescaling $(1+\alpha )s\rightarrow s$,
$\tilde{s}\rightarrow\tilde{s}\rightarrow(1+\alpha)\tilde {s}$, the reduced
response functional takes on the form
\begin{align}
\mathcal{J}_{red} &  =\int d^{d}x\Big\{\int_{0}^{\infty}dt\,\lambda\tilde
{s}(t)\bigl[\lambda^{-1}\dot{s}(t)-k\nonumber\\
&  -\int_{0}^{s(t)}ds(t^{\prime})\,D\bigl(s(t)-s(t^{\prime
})\bigr)\nabla^{2}s(t^{\prime})\bigr]\nonumber\\
&  -\frac{\lambda^{2}f}{2}\int_{0}^{\infty}dtdt^{\prime}\tilde{s}%
(t)C\bigl(s(t)-s(t^{\prime})\bigr)\tilde{s}(t^{\prime}%
)\Big\}\,.\label{J-red-1}%
\end{align}
A more detailed account of the steps leading from Eq.~(\ref{J-red}) to (\ref{J-red-1}) can be found in Appendix~\ref{appenDeJ}.

Note that this is the dynamic response functional of the qEW model with an
additional retardation of the elastic term $\sim\nabla^{2}s$ described by the
function
\begin{equation}
D(s)=\delta(s)-(1-g/f)G(s)\,.
\end{equation}
The corresponding Langevin equation reads as follows%
\begin{align}
&
\lambda^{-1}\dot{s}(\mathbf{x},t)=
\int_{0}^{s(\mathbf{x},t)}ds(\mathbf{x},t^{\prime})\,D\bigl(s(\mathbf{x},t)
-s(\mathbf{x},t^{\prime})\bigr)\nabla^{2}s(\mathbf{x},t^{\prime})\nonumber\\
& \qquad\qquad\qquad\qquad\qquad\qquad+k+\mathcal{F}(\mathbf{x},t)\,,\nonumber\\
& \langle
\mathcal{F}(\mathbf{x},t)\mathcal{F}(\mathbf{x}^{\prime},t^{\prime})\rangle
=fC\bigl(s(\mathbf{x},t)-s(\mathbf{x},t^{\prime})\bigr)\,\delta(\mathbf{x}%
-\mathbf{x}^{\prime})\,.
\end{align}

Next, we will show that the deviation of the ratio $g/f$ from $1$ does not
lead to new renormalizations, and is therefore a redundant, inessential
parameter. We consider the cumulant-generation functional%
\begin{equation}
\mathcal{W}[\tilde{h},h]=\ln\Big\{\int\mathcal{D}(\tilde{s},s)\exp
\bigl(-\mathcal{J}_{red}[\tilde{s},s]+(\tilde{h},\tilde{s}%
)+(h,s)\bigr)\Big\}\,,
\end{equation}
where $(h,s)$ denotes the integral of the functions $h$ and $s$ over space-time,
and use the so-called statistical tilt symmetry invariance $s(\mathbf{x}%
,t)\rightarrow s(\mathbf{x},t)+v(\mathbf{x})$. Exploiting this invariance, it
is easy to see that the cumulant-generation functional $\mathcal{W}$ has the
property%
\begin{equation}
\mathcal{W}\bigl[\tilde{h}+\lambda\bar{D}\nabla^{2}v,h\bigr]+(h,v)=\mathcal{W}%
\bigl[\tilde{h},h\bigr]\,,
\end{equation}
where $\bar{D}=\int_{0}^{\infty}ds\,D(s)=g/f$. A functional derivative with
respect to $v(\mathbf{x})$ produces
\begin{equation}
\int dt\Big(\lambda\bar{D}\nabla^{2}\frac{\delta\mathcal{W}\bigl[\tilde{h}%
,h\bigr]}{\delta\tilde{h}(\mathbf{x},t)}+h(\mathbf{x},t)\Big)=0\,.
\end{equation}
Taking a further functional derivative with respect to $h(\mathbf{x}^{\prime
},t^{\prime})$, we obtain
\begin{equation}
\bar{D}\nabla^{2}\int
dt\lambda\frac{\delta\mathcal{W}\bigl[\tilde{h},h\bigr]}{\delta
h(\mathbf{x}^{\prime},t^{\prime})\delta\tilde{h}(\mathbf{x},t)}+\delta
(\mathbf{x-x}^{\prime})=0\,.
\end{equation}
For the Fourier transform of the full response function
\begin{align}
R(\mathbf{x},t)  &  =\int_{\mathbf{q},\omega}R_{\mathbf{q},\omega}
\,\mathit{e}^{i(\omega t-\mathbf{q\cdot x})}=\lambda\langle
s(\mathbf{x},t)\tilde
{s}(0,0)\rangle\nonumber\\
&  =\lambda\left.  \frac{\delta W\bigl[\tilde{h},h\bigr]}{\delta h(\mathbf{x}%
,t)\delta\tilde{h}(\mathbf{0},0)}\right\vert _{\tilde{h}=h=0}\,,
\end{align}
this leads to
\begin{equation}
R_{\mathbf{q},\omega=0}=\frac{1}{\bar{D}\,q^{2}}\,.
\end{equation}
Hence, the full static response function does not acquire any additional
contributions in perturbation theory, and consequently $\bar{D}$ does not need
any renormalization. Therefore the function
$D(s)$ can be safely approximated by a delta-function $\delta(s)$, a step
which merely leads to a resetting of the time-scale by
$\lambda\rightarrow\lambda f/g$ as correctly observed by LeDW. In other words, the effects of the retardation term can be transformed away, and  therefore the latter does not contribute to the critical behavior. According to Wegner's canonical classification scheme of field theoretical operators \cite{Weg74} (see also, e.g.,
Ref.~\cite{ZinnJust89}), such a term is called redundant.

By letting $D(s) \to \delta(s)$, we untinately
obtain from the reduced dynamic response functional (\ref{J-red-1}) the
well-known depinning- or
qEW-functional%
\begin{align}
\mathcal{J}_{qEW}  &  =\int d^{d}x\Big\{\int_{0}^{\infty}dt\,\lambda\tilde
{s}(t)\bigl[\lambda^{-1}\dot{s}(t)-\nabla^{2}s(t)-k\bigr]\nonumber\\
&  -\frac{\lambda^{2}}{2}\int_{0}^{\infty}dtdt^{\prime}\tilde{s}%
(t)\Delta\bigl(s(t)-s(t^{\prime})\bigr)\tilde{s}(t^{\prime})\Big\}\,,
\end{align}
with the starting noise correlation%
\begin{equation}
\Delta(s)=\frac{f^{3}}{2g^{2}}\exp(-\left\vert s\right\vert )\,,
\end{equation}
Note that this noise correlation shows the characteristic kink for $s\rightarrow0$.

Finally, we would like to comment on the remaining terms that arise from the retardation of the
elastic term after shrinking it to a $\delta$-function. To study the behavior of the leading such term under the RG, one should
determine the renormalization of insertions of a time-bilocal operator of the form%
\begin{equation}
\mathcal{O}(\mathbf{x},t,t^{\prime})=\tilde{s}(\mathbf{x}%
,t)O\bigl(u(\mathbf{x},t,t^{\prime})\bigr)\frac{\partial u(\mathbf{x}%
,t,t^{\prime})}{\partial t^{\prime}}\nabla^{2}u(\mathbf{x},t,t^{\prime})
\end{equation}
where $u(\mathbf{x},t,t^{\prime})=s(\mathbf{x},t)-s(\mathbf{x},t^{\prime})$
with $O(u)$ denoting some function of $u$. Note that if the qEW-model in its original form
is renormalizable -- as it is generally assumed -- such operator-insertions have to be irrelevant.

\section{Concluding remarks}

In summary, we have taken a fresh look at the derivation of field theoretic
functional describing C-DP and the recent mapping of C-DP onto qEW.

Our derivation of the C-DP dynamic response functional utilizes the
Grassberger-transformation. When it can be applied, the
Grassberger-transformation has tangible advantages over the Doi-shift. The
field variables produced by the Grassberger-transformation correspond to actual
particle densities and their conjugate response field, and we think that they
are more natural and more intuitive than the field variables induced by the
Doi-shift. The Grassberger-transformation is particularly useful for systems
where typical particle configurations correspond to agglomerations of fractal
clusters, as is the case for C-DP. Then the scaling dimensions of the density
and response fields are typically positive, and terms of higher than harmonic
order in the response field are usually irrelevant so that it is justified to
truncate at second order the expansion of the exponentials arising arising
through the transformation. A further advantage is that, if
particle-conservation holds as for C-DP, the corresponding symmetry of the
response functional is realized linearly. Part of our motivation for the
present paper is to highlight the usefulness of the Grassberger-transformation
as the concise method to capture the emergent universal description of
reaction-diffusion systems in form of a coarse-grained effective stochastic
equation of motion, and we hope that its usefulness will be appreciated more in
the future. For the reader who is interested in a deeper discussion of the
applicability of the Grassberger-transformation to coherent state path
integrals in general, we have compiled some additional thoughts in 
Appendix~\ref{appenA}.

The profound work by LeDW essentially settled a long-standing issue in
statistical physics by mapping C-DP onto qEW. The full response functional for
C-DP originally contains an additional term, which is superficially relevant
below the upper critical dimension $d_c=4$, and if this term had to be
retained, the direct mapping would fail. We show rigorously that this term,
which generates a retardation of the elastic interaction in the qEW, does not
produce any additional contributions to the static response function at any
order in perturbation theory. Thus, the retardation term does not flow under
the action of the RG. Consequently, it does not impact the asymptotic scaling
behavior of C-DP and therefore can be dropped from the response functional.
Note that the retardation term, since it does not flow, it does
not flow to zero, in particular, as an irrelevant term would. There are perhaps
different interpretations of the notion of irrelevance in the context of the
RG. In our personal opinion, the cleanest interpretation or definition is the
one given in Wegner's classification scheme, and according to this classification, the retardation
term is redundant. Irrelevant and redundant terms, respectively,
behave qualitatively differently under the action of the RG, however,
ultimatively both can and should be discarded from minimal field theoretic
models. Fortunately, after the dust settles, the omission of the retardation
term is justified, albeit on different grounds, and the outcome of the LeDW
mapping stands correct.

Finally, we would like to point out that the appearance of the retardation term in the qEW functional is not merely a consequence of the mapping from C-DP but that this term is inherent in the qEW model itself. When one studies fluctuation effects based on the original qEW functional that has no retardation term (except for noise contributions), diagrammatic perturbation theory does eventually produce such a term, and this term is exactly of the form discussed above. The qEW model is generally accepted as being renormalizable as it stands. One can show that the retardation term, though marginal on dimensional grounds, does not require an independent renormalization as a consequence of statistical tilt-symmetry, and hence it is redundant and can be omitted. In other words, the redundant retardation term is native to qEW.

\begin{acknowledgments}
This work was supported by the NSF under No.~DMR-1104701 (OS) and
No.~DMR-1120901 (OS).
\end{acknowledgments}

\appendix

\section{Doi-shift versus Grassberger-transformation}
\label{appenA}

It is well known that particle-reactions and -diffusion modeled by
master-equations, and described by \textquotedblleft
second-quantized\textquotedblright\ Fock-space methods, can be conveniently
formulated as coherent-state path-integrals (CSPI)
\cite{Doi76,GraSche80,Pel84,Uwe14,TaHoVo05,Wie15}. For simplicity, we consider
here
only one sort of particles, $A$, with reactions%
\begin{equation}
kA\overset{r_{k,l}}{\longrightarrow}lA
\end{equation}
and reaction-rates $r_{k,l}$. After applying a (naive) continuum limit, the
corresponding CSPI-action with an additional diffusion-term is given by%
\begin{align}
\mathcal{S} &  =\int d^{d}x\int_{-\infty}^{\infty}dt\Big\{(\hat{a}%
-1)\partial_{t}a+\lambda\nabla\hat{a}\cdot\nabla a\nonumber\\
&  -\sum_{k,l}r_{k,l}\big(\hat{a}^{l}-\hat{a}^{k}\big)a^{k}%
\Big\}\,.\label{action}%
\end{align}
Here, the fields $\hat{a}$ and $a$ correspond to the coherent-state
eigenvalues of the bosonic creation and annihilation operators of the
Fock-space with initial and final conditions $\hat{a}(\infty)=1$,
$a(-\infty)=0$. It is important to understand that the complex field $a$ is
not the particle-density which is given by $n=\hat{a}a$, a real semipositive
quantity. Following Doi, usually a field shift according to $\hat{a}%
=1+\tilde{a}$ is applied. This Doi-shift results in%
\begin{align}
\mathcal{S} &  =\int d^{d}x\int_{-\infty}^{\infty}dt\Big\{\tilde{a}%
\partial_{t}a+\lambda\nabla\tilde{a}\cdot\nabla a-\tilde{a}\sum_{k,l}%
(l-k)r_{k,l}a^{k}\nonumber\\
&  -\frac{\hat{a}^{2}}{2}\sum_{k,l}(l+k-1)(l-k)r_{k,l}a^{k}+\ldots\Big\}\,,
\end{align}
where the series stops with the quadratic term $\hat{a}^{2}$ if $l,k\leq2$. In
a corresponding Langevin description, this quadratic term is often interpreted
as a noise term. Then one gets the well known result: branching ($l>k$) leads
to real noise whereas annihilation ($l<k$ with exception of $k=1$) leads to
imaginary noise which is to interpreted as a first passage problem
\cite{Wie15}. This type of behavior is characteristic for processes where
random walkers sparsely distributed in space which meet and react from time to
time, however, it is not so for systems of clusters of, in general, fractal
particle-agglomerations (typical for percolating processes). For the latter
type of systems, it is more appropriate to switch to the density $n$ as the
fundamental variable. This is achived by Grassbergers quasi-canonical
transformation~\cite{Gra82,Ja01b,ABBL06}
\begin{equation}
a=n\exp(-\tilde{n})\,,\quad\hat{a}=\exp(\tilde{n})\,,\label{Gra-Tr}%
\end{equation}
which is similar to an inverse Cole-Hopf transformation. Note that a creation
of a state with $\rho_{i}$ particles in cell $i$ of a spatially distributed
system corresponds in the CSPI to an insertion of the product%
\begin{equation}%
%TCIMACRO{\dprod \limits_{i}}%
%BeginExpansion
{\displaystyle\prod\limits_{i}}
%EndExpansion
\hat{a}_{i}^{\rho_{i}}=\exp\Big(\sum_{i}\rho_{i}\tilde{n}_{i}\Big)\rightarrow
\exp\Big(\int d^{d}x\,\rho(\mathbf{x})\tilde{n}(\mathbf{x}%
)\Big)\,,\label{particle insertion}%
\end{equation}
in the continuum limit. Hence, coarse-graining is simply performed by
neglecting higher Fourier components of $\tilde{n}$. We will show in a moment
that this product can be simply interpreted as a creation process%
\begin{equation}
0\overset{\rho}{\longrightarrow}A\label{particle creation}%
\end{equation}
in the transformed action (\ref{action}) $\mathcal{S}\left[  \hat{a},a\right]
\rightarrow\mathcal{J}\left[  \tilde{n},n\right]  $ of the CSPI.

Applying the Grassberger-transformation (\ref{Gra-Tr}), we obtain after some
partial integrations using $\tilde{n}(\infty)=n(-\infty)=0$%
\begin{align}
\mathcal{J}\left[  \tilde{n},n\right]   &  =\int d^{d}x\int_{-\infty}^{\infty
}dt\Big\{\tilde{n}\partial_{t}n+\lambda\nabla\tilde{n}\cdot\nabla n-\lambda
n\left(  \nabla\tilde{n}\right)  ^{2}\nonumber\\
&  -\sum_{k,l}r_{k,l}\big(\exp\bigl((l-k)\tilde{n}\bigr)-1\big)n^{k}%
\Big\}\nonumber\\
&  =\int d^{d}x\int_{-\infty}^{\infty}dt\Big\{\tilde{n}\partial_{t}%
n+\lambda\nabla\tilde{n}\cdot\nabla n-\lambda n\left(  \nabla\tilde{n}\right)
^{2}\nonumber\\
&  -\tilde{n}\sum_{k,l}(l-k)r_{k,l}n^{k}-\frac{\tilde{n}^{2}}{2}\sum
_{k,l}(l-k)^{2}r_{k,l}n^{k}\nonumber\\
&  +\ldots\Big\}\,.\label{respfunct}%
\end{align}
Note that the particle insertion (\ref{particle insertion}) into the
path-integral corresponds to a particle-creation process
(\ref{particle creation}) that leads to a term with $k=0$ and $l=1$ in the
functional (\ref{respfunct}). The expansion of the exponential in
$\mathcal{J}$ is analogous to a Kramers-Moyal expansion of the
master-equation. Skipping all the terms higher than the quadratic term
$\sim\tilde{n}^{2}$ leads to the well-known dynamic-response functional
\cite{Ja76,DeDo76,Ja92} of reaction-diffusion systems represented by a
Langevin equation (in Ito-interpretation)%
\begin{equation}
\partial_{t}n=\lambda\nabla^{2}n+R(n)+\zeta\,,
\end{equation}
where the rate is $R(n)=\sum_{k,l}(l-k)r_{k,l}n^{k}=:\sum_{k}r_{k}n^{k}$, and
$\zeta$ is a real Gaussian noise with correlator%
\begin{equation}
\langle\zeta(\mathbf{x},t)\zeta(\mathbf{x}^{\prime},t^{\prime})\rangle
=\bigl[Q(n)-\lambda\nabla n\nabla\bigr]\delta(\mathbf{x}-\mathbf{x}^{\prime
})\delta(t-t^{\prime})
\end{equation}
with $Q(n)=\sum_{k,l}(l-k)^{2}r_{k,l}n^{k}=:\sum_{k}q_{k}n^{k}$.

The truncation of the expansion at second order deserves some further
scrutiny~\cite{TaHoVo05}. Let us carefully examine the behavior of the
expansion under coarse graining. To this end, we asses the scaling behavior of
the fields near the upper dimension above which a simple mean-field
approximation is correct. Naively, the response field $\tilde{n}$ is
dimensionless, and the particle-density scales as $n\sim\mu^{d}$ where $\mu$ is
an inverse length scale. Thus, the reaction-constants scale naively as
$r_{k}\sim q_{k}\sim\lambda\mu^{2-(k-1)d}$. Let $k_{0}$ be the lowest $k$ so
that $r_{k}\neq0$, and $k_{1}$ the lowest $k$ with $q_{k}\neq0$. Then
$k_{0}\geq k_{1}$, and a rescaling of the fields with $\mu$ so that
$\tilde{n}n^{k_{0}}\sim\tilde{n} ^{2}n^{k_{1}}$ and $\tilde{n}n\sim\mu^{d}$
leads to
\begin{equation}
\tilde{n}\sim\mu^{d(k_{0}-k_{1})/(k_{0}-k_{1}+1)}\,,\quad
n\sim\mu^{d/(k_{0}-k_{1}+1)}\,.
\end{equation}
Thus, if $k_{0}>k_{1}$ ($k_{0}=k_{1}+1$ in typical cases) both fields receive
positive scaling dimensions ($\tilde{n}\sim n\sim\mu^{d/2}$ if
$k_{0}=k_{1}+1$), and all higher order monomials in $R(n)$ and $Q(n)$ as well
as all contributions proportional to $\tilde{n}^{l}$ with exponents $l>2$ are
irrelevant and contribute only corrections to the leading scaling behavior
generated by the relevant leading terms $\tilde{n}n^{k_{0}}$ and
$\tilde{n}^{2}n^{k_{1}}$. Even the diffusional noise is sub-leading if
$k_{1}<2$. This reasoning justifies the truncation of the expansion at second
order. It shows that the application of the Grassberger-transformation in such
cases is the concise method to capture the emergent universal description of
reaction-diffusion systems.

However, if $k_{0}=k_{1}$, typical for annihilation reactions, the
Grassberger-transformation to density fields is not applicable, and one has to
deal with the original formulation of the CSPI. The superficial scaling
behavior is correct, of course, above and at a critical dimension $d_{c}$, and
below $d_{c}$, perturbational corrections appear. The critical dimension is
determined by the similarity $\tilde{n}n^{k_{0}}\sim\tilde{n}^{2}n^{k_{1}}\sim
\mu^{d_{c}+2}$. One obtains
\begin{equation}
d_{c}=2\frac{(k_{0}-k_{1}+1)}{(k_{0}-1)}\,.
\end{equation}

\section{Doi-shift, truncations, imaginary and real diffusional noise}
\label{appenTrunc}
Here, we present an alternate derivation of the dynamical response functional, Eq.~(\ref{J-functional}), that uses the Doi-shift instead of the Grassberger-transformation. Our motivation here is twofold. First, we think that is is instructive to see the two approaches side by side so that one can compare them in a specific example. Most readers will agree with us that the route via the Doi-shift is significantly less intuitive and more cumbersome than the one taken in the main text. Second, we feel that the way the derivation of the original, full (with diffusion of the A-particles) C-DP functional has been presented in the literature might leave the reader wondering where some of the key terms come from. Thus, we review here some of the essential steps involved in the derivation.

Our starting point is the CSPI-action, Eq.~(\ref{CSPI-act}), augmented by terms that arise when the diffusion of A-particles is permitted~\cite{WiOeHi98}:
\begin{align}
\mathcal{S}= &  \int d^{d}xdt\Big\{(\hat{a}-1)\partial_{t}a+(\hat
{b}-1)\partial_{t}b+D\nabla\hat{b}\cdot\nabla b+D^{\prime}\nabla\hat{a}%
\cdot\nabla a\nonumber\\
&  -k_{1}\big(\hat{b}^{2}-\hat{a}\hat{b}\big)ab-k_{2}\big(\hat{a}\hat{b}%
-\hat{b}^{2}\big)b^{2}-k_{3}\big(\hat{a}-\hat{b}\big)b\Big\}\nonumber\\
= &  \int d^{d}xdt\Big\{\tilde{a}\partial_{t}a+\tilde{b}\partial_{t}%
b+D\nabla\tilde{b}\cdot\nabla b+D^{\prime}\nabla\tilde{a}\cdot\nabla
a\nonumber\\
&  +(\tilde{b}-\tilde{a})\bigl[k_{3}+k_{2}b-k_{1}a\bigr]b-(\tilde{b}-\tilde
{a})\tilde{b}\bigl[k_{1}a-k_{2}b\bigr]b\Big\}\,.
\end{align}
Now, let's focus on the noise term (the one of second order in the fields with the tilde). In this term, we truncated as follows,
\begin{equation}
k_{1}a-k_{2}b=k_{1}c_{0}+\ldots\,,
\end{equation}
where $c_{0}$ is the constant initial value of $c$. Setting $\tilde{b}^{\prime}=\tilde
{b}-\tilde{a}$, the truncated noise term becomes
\begin{align}
&-(\tilde{b}^{\prime2}+\tilde{a}\tilde{b}^{\prime})bk_{1}c_{0}=-(\tilde
{b}^{\prime2}+\tilde{a}\tilde{b}^{\prime})
\nonumber\\
&=-%
\begin{pmatrix}
\tilde{b}^{\prime}, & \tilde{a}%
\end{pmatrix}%
\begin{pmatrix}
1 & 1/2\\
1/2 & 0
\end{pmatrix}%
\begin{pmatrix}
\tilde{b}^{\prime}\\
\tilde{a}%
\end{pmatrix}
bk_{1}c_{0}\,.
\end{align}
Note that the matrix on the right hand side has a positive and a negative eigenvalue implying that the noise described by this noise term has both real and imaginary components as recently pointed out by Wiese~\cite{Wie15}. Note also that the problematic term $\tilde{a}\tilde{b}^{\prime}$ that originally appeared in Ref.~\cite{WiOeHi98} was absent in the follow-up paper~\cite{OeWijLerHi00}, where a real diffusional noise term appeared instead. This brings up the question of how imaginary noise can become real diffusional noise.

To answer this question, let us substitute
\begin{equation}
a^{\prime}=a+c_{0}\tilde{a}
\end{equation}
and integrate out $\tilde{a}\partial_{t}\tilde{a}=1/2\,\partial_{t}\tilde{a}^{2}$. Then the CSPI action becomes
\begin{align}
\mathcal{S}&=\int d^{d}xdt    \Big\{\tilde{a}\partial_{t}(a^{\prime}%
+b)+\tilde{b}^{\prime}\partial_{t}b+D\nabla\tilde{b}^{\prime}\cdot\nabla b
\nonumber\\
&+\nabla\tilde{a}\cdot(D\nabla b+D^{\prime}\nabla a^{\prime})+\tilde{b}^{\prime}\bigl[k_{3}+k_{2}b-k_{1}a^{\prime}\bigr]b
\nonumber\\
&  -k_{1}%
c_{0}\,\tilde{b}^{\prime2}b-D^{\prime}c_{0}\,(\nabla\tilde{a})^{2}\Big\}\,,
\end{align}
where the real diffusional noise has replaced the imaginary noise.

To proceed from here to our dynamical response functional $\mathcal{J}$, we set  $a^{\prime}+b=c$, $\tilde{a}=\tilde{c}$, $b=n$, $\tilde
{b}^{\prime}=n$ and $D^{\prime}=0$. Up to a trivial redefinition of coefficients and the source terms that we still need to add, this produces $\mathcal{J}$ as given in Eq.~(\ref{J-functional}).

Finally, let us compile what the switch from $a$ to $a^{\prime}$ means on the level of the particle densities $n_A$ and $n_B$ and the total particle density $c$. Expansion, truncation and this switch give
\begin{subequations}
\begin{align}
n_{A} &  =\mathit{e}^{\tilde{n}_{A}}a=a+\tilde{a}a+\ldots\\
&  =a+\tilde{a}\,(c_{0}+\ldots)+\ldots=a^{\prime}+\ldots\,,\\
n_{B} &  =\mathit{e}^{\tilde{n}_{B}}b=b+\ldots\,,\\
c &  =n_{A}+n_{B}=a^{\prime}+b+\ldots\,,
\end{align}
\end{subequations}
in full agreement with the approach based on the Grassberger-transformation.

As pointed out earlier, we feel that the approach based on the Grassberger-transformation is much clearer and more elegant than the one based on the Doi-shift. After all, the essential physical ingredient of the model is that total particle density fluctuates about its fixed finite initial value $c_{0}$, and the resulting theory is based on an expansion in the fluctuations of the density fields $c$, $n$ and their corresponding response fields.  The natural advantage of the Grassberger-transformation hereby is that it describes physical densities from the onset.

\section{Derivation of the reduced response functional $\mathcal{J}_{red}$}
\label{appenDeJ}

In this Appendix, we provide some details on how to proceed from the reduced response functional $\mathcal{J}_{red}$ as stated in Eq.~(\ref{J-red}) to its form stated in Eq.~(\ref{J-red-1}). We focus on the most difficult term, i.e., the last term in these two equations. We assume the usual rules of Ito-calculus, in particular
\begin{equation}
\theta(s)=\theta(s+0)\,,\quad\delta(s)=\delta(s+0)\,.
\end{equation}

Applying Laplace-transformation to the propagator given in Eq.~(\ref{propAndCorr}), we have
\begin{equation}
\hat{G}(z)=\int_{0}^{\infty}ds\,\mathrm{e}^{-zs}G(s)=\frac{1}{z+1}\,.
\end{equation}
For the Laplace-transformation of the kernel $B(s)$ appearing in the equation of motion, Eq.~(\ref{Bewegl}), this implies
\begin{equation}
\hat{B}(z)=1+\frac{\alpha}{z+1}=\frac{z+1+\alpha}{z+1}\,.
\end{equation}
Because Laplace-transformation reduces convolution integrals to simple multiplications, it produces
\begin{equation}
\hat{K}(z)=\frac{z+1}{z+1+\alpha}=1-\frac{\alpha}{z+1+\alpha}
\end{equation}
for the inverse of the kernel $B(s)$. Note that this is nothing but the Laplace-transformation of Eq.~(\ref{D(s)}).

Now, we are in position to apply Laplace-transformation to the kernel $C(s)$ in Eq.~(\ref{J-red}). Writing the correlator as
\begin{equation}
C(s-s^{\prime})=\int ds_{0}G(s-s_{0})G(s^{\prime}-s_{0})\,,
\end{equation}
we have
\begin{align}
\label{kernalCProd}
&g\int\int dtdt^{\prime}\tilde{s}(t)\tilde{s}(t^{\prime})C(s(t)-s(t^{\prime
}))
\nonumber\\
&=g\int ds_{0}\Big(\int dt\,\tilde{s}(t)G(s(t)-s_{0})\Big)
\nonumber\\
&\times\Big(\int
dt^{\prime}\,\tilde{s}(t^{\prime})G(s(t^{\prime})-s_{0})\Big)\,.
\end{align}
Under the transformation (\ref{sTildeTrans}), the individual factors on the right hand side of Eq.~(\ref{kernalCProd}) transform as
\begin{align}
&\int dt\,\tilde{s}(t)G(s(t)-s_{0})
\nonumber\\
&  \rightarrow\int dt^{\prime}\tilde
{s}(t^{\prime})\int dt\,K(s(t^{\prime})-s(t))\dot{s}(t)G(s(t)-s_{0}%
)\nonumber\\
&  =\int dt^{\prime}\tilde{s}(t^{\prime})\int ds\,K(s(t^{\prime}%
)-s)G(s-s_{0})\,.
\end{align}
Upon Laplace-transformation, the convolution of $K$ und $G$ turns into a product of $\hat{K}$ und $\hat{G}$:
\begin{equation}
\hat{K}(z)\hat{G}(z)=\frac{1}{z+1+\alpha}\,.
\end{equation}
This is the same as the result of Laplace-transformation applied to $G((1+\alpha)s)$. Thus, we have
\begin{align}
&g\int\int dtdt^{\prime}\tilde{s}(t)\tilde{s}(t^{\prime})C(s(t)-s(t^{\prime
}))
\nonumber\\
&\rightarrow\frac{g}{1+\alpha}\int\int dtdt^{\prime}\tilde{s}(t)\tilde
{s}(t^{\prime})C\bigl((1+\alpha)(s(t)-s(t^{\prime}))\bigr)\,.
\end{align}
Finally, applying the rescaling $(1+\alpha)s\rightarrow s$, $\tilde{s}\rightarrow(1+\alpha)\tilde{s}$ and setting $(1+\alpha)=f/g$  completes the journey from Eq.~(\ref{J-red}) to  Eq.~(\ref{J-red-1}).


\begin{thebibliography}{99}
\bibitem{footnote}The fact that $(\dot{s})^2$ appears in the second line of Eq.~(\ref{J-red}) can be easily understood by moving the last term in Eq.~(\ref{Jagain}) into the first line of this equation where it becomes $(\dot{s})^2$ instead of $\dot{s}$.
\end{thebibliography}
\end{document}